# Prepare for Trouble and Make it Double! Supervised – Unsupervised Stacking for Anomaly-Based Intrusion Detection

*Tommaso Zoppi\*, Andrea Ceccarelli*

*Department of Mathematics and Informatics, University of Florence,Viale Morgagni 65, 50142 - Florence - Italy*

A B S T R A C T

In the last decades, researchers, practitioners and companies struggled in devising mechanisms to detect malicious activities originating security threats. Amongst the many solutions, network intrusion detection emerged as one of the most popular to analyze network traffic and detect ongoing intrusions based on rules or by means of Machine Learners (MLs), which process such traffic and learn a model to suspect intrusions. Supervised MLs are very effective in detecting known threats, but struggle in identifying zero-day attacks (unknown during learning phase), which instead can be detected through unsupervised MLs. Consequently, supervised and unsupervised MLs have their own advantages and downfalls that complement each other. Unfortunately, there are no definitive answers on the combined use of both approaches for network intrusion detection. In this paper we first expand the problem of zero-day attacks and motivate the need to combine supervised and unsupervised algorithms. We propose the adoption of meta-learning, in the form of a two-layer Stacker, to create a mixed approach that detects both known and unknown threats. Then we implement and empirically evaluate our Stacker through an experimental campaign that allows i) debating on meta-features crafted through unsupervised base-level learners, ii) electing the most promising supervised meta-level classifiers, and iii) benchmarking classification scores of the Stacker with respect to supervised and unsupervised classifiers. Last, we compare our solution with existing works from the recent literature. Overall, our Stacker reduces misclassifications with respect to (un)supervised ML algorithms in all the 7 public datasets we considered, and outperforms existing studies in 6 out of those 7 datasets. In particular, it turns out to be more effective in detecting zero-day attacks than supervised algorithms, limiting their main weakness but still maintaining adequate capabilities in detecting known attacks.

Available at the Elsevier JNCA

## 1   Introduction

Nowadays ICT systems often embed functionalities that are partially realized, controlled or monitored by distributed sub-systems. Paradigms as Cyber-Physical Systems (CPSs), Cloud Systems and even Cluster Computing usually network together different components or subsystems to deliver such functionalities. To this extent, they often offer remote access to services and relevant information, and they may consequently be subject to (cyber)attacks [2], [6], [92], [95]. In the last decade, such cyber-threats had a constantly growing impact as pointed out by several reports [7], [8]. Consequently, *Intrusion Detection Systems* (IDSs) [3], [4] became critical building blocks to detect potential threats and trigger modules that are able to block or mitigate the adverse effects of cyber-threats. IDSs collect and analyse data from networks and - often - system indicators to potentially detect malicious or unauthorized activities, based on the hypothesis that an ongoing attack has distinguishable effects on such indicators.

### 1.1   Signatures and Anomalies

Network data and system indicators are used since decades as baseline to derive *signatures* of known attacks. For example, malformed headers in packets or emails, that may generate problems as resource exhaustion [9], can be detected by range-checking fields of the packets that arrive from the network. Moreover, ransomwares can be effectively detected [10] by analysing system calls to check if they compute cryptographic functions. Signature-based approaches perform satisfactorily when aiming at known attacks [2], [3], [4]; on the other hand, they exhibit weaknesses in detecting variations of known attacks or *zero-day attacks* [6], whose signature is unknown.

\* *Corresponding author.* Tel.: +39 055 2751483; E-mail address: tommaso.zoppi@unifi.it

The alternative to signature-based intrusion detection is offered by *anomaly-based* [1] intrusion detection. Suspicious activities can be detected through Machine Learning (ML) algorithms that rely on past network and system observations to learn a model that classifies novel data points into either normal data points or anomalous data points i.e., potential attacks. Figure 1 shows how a (binary) classifier is first trained (Tr1, Tr2) to learn a model. Training is mainly performed using labels, which indicate if data points correspond to attacks or not. Then, the classifier predicts the label for a novel data point (Te1) by running a ML algorithm which outputs a numerical value (e.g., it may compute the distance to the centroid of the nearest cluster, Te2). This value is sent to a decision function (e.g.: "is the data point sufficiently close to the centroid of the cluster, such that we can assume the data point belongs to the cluster?", Te3): the decision function outputs a binary label (Te4) which decides if the data point is anomalous or not. The decision function may be embedded into the classifier, or be a parameter of the classifier itself. In our clustering example, it may i) be a static threshold, ii) depend on the sparsity of the cluster, or iii) be related to the amount of items in the cluster, and so on.

## 1.2 Supervised and Unsupervised Intrusion Detection

Intrusion Detectors that rely on *Supervised* ML algorithms require historical system observations for which the label (also called *class*) is known. They learn a model to classify any new observation – data point – either as collected i) when a system is under malicious attack, or ii) during normal operations. For example, the literature reports on the successful usage of Random Forests [31], Support Vector Machines [33], Convolutional [34] and Deep [35], [38] Neural Networks for the detection of security threats through the analysis of network traffic. These supervised ML algorithms are very effective in detecting known threats, albeit they cannot effectively deal with novel or unknown threats. Unfortunately, this is a severe weakness as many threats cannot be identified at system design time.

On the other hand, *unsupervised* ML algorithms do not assume any knowledge on the attacks. They model the expected (normal) behaviour of the system, and classify any deviation from the normal behaviour as anomaly, i.e., suspected attacks [1]. Clustering algorithms [65], [69] are probably the most widespread unsupervised ML algorithms, despite statistical [60], angle [46], density [70], [63] algorithms, and unsupervised variants of neural networks [64], neighbour-based [61], [71] or classification [72], [62] algorithms were proven to be valid alternatives [29], [30]. Since unsupervised ML algorithms build their model without relying on labels, they do not distinguish between known and unknown or zero-day attacks. However, they are more prone to misclassifications of known attacks than supervised algorithms.

## 1.3 On the Complementarity and Combination of Supervised and Unsupervised Algorithms

As evident from the above discussion, supervised and unsupervised approaches are complementary.

Supervised techniques learn from past attack episodes, and mainly model the (known) malicious behaviours. In fact, the adoption of supervised classifiers always carries a deliberate weakness in detecting zero-day threats, or (slight) modifications of known attacks. While it is true that classifiers may employ techniques to prevent overfitting (e.g., pruning Decision Trees [5]) and therefore avoid building a model that corresponds too closely to a particular attack, *there is no reliable way to make supervised classifiers suitable to detect previously unknown threats*. Instead, unsupervised techniques model a normal behaviour and classify any deviation as anomaly.

Each holds its own strengths and weaknesses, and as such they perform differently depending on the scenario; as a consequence, it is highly desirable to pair supervised classifiers with unsupervised classifiers as in [12], [13], [14], [17]. Indeed, most of these solutions are very problem-specific and may hardly be generalized. Additionally, combining both approaches is not trivial and does not always result in improved capabilities: some misleading algorithms may let the combination of both approaches lean towards a misclassification, with obvious detrimental effects.

Potential strategies to tackle these challenges come from the *Meta-Learning* [74] domain. A Meta-Learner (or Meta-Classifier) orchestrates ensembles of MLs to improve classification capabilities. For example, well-known supervised classifiers as Random Forests [31] and ADABoost [32] respectively build homogeneous *Bagging* [21] and *Boosting* [22] ensembles of decision trees to compute the final result. Instead, *Voting* [36] is widely used by system architects to design and implement redundant systems [37]. Other approaches as *Stacking* [23], *Cascading* [24], *Arbitrating* [27] and *Delegating* [26] are less notorious but still deemed relevant [28].

## 1.4 Contribution and Paper Structure

To the best of our knowledge, in the literature there is neither clear agreement nor experimental confirmation on the proper approach to orchestrate and combine supervised and unsupervised classifiers to detect network intrusions, despite several tries have been made throughout years. Consequently, this paper first recaps on available strategies to combine classifiers for attack detection (Section 2) and on suitability of Meta-Learning in combining or merging both supervised and unsupervised classifiers (Section 3). Such discussion lets Section 4 proposing an approach based on Stacking with multiple unsupervised base-level learners and a supervised meta-level learner to improve the detection of both known and unknown - zero-day - threats.

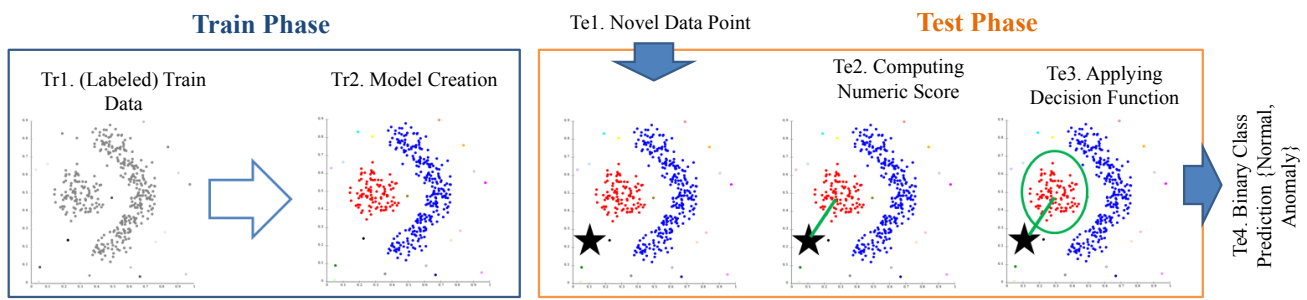

Figure 1: Main Steps for Train (Tr1, Tr2) and Test (Te1, Te2, Te3, Te4) phases of Binary Classifiers.

Then, we plan, setup and execute in Section 5 an experimental campaign using public attack datasets and well-known unsupervised/supervised algorithms that aims to show the advantages of our Stacker over state-of-the-art algorithms. This allows Section 6 debating on how such base-learners provide additional and relevant meta-features [75], paving the way for discussing the classification performance of our approach with respect to supervised and unsupervised algorithms. It turns out that our Stacker classifies known attacks better than supervised algorithms while noticeably enhancing the detection of zero-day attacks. Then, Section 7 shows how to build a Stacker for a target system, letting Section 8 to compare results with other studies in the literature that report quantitative metrics on intrusion detection. Lastly, Section 9 concludes the paper and elaborates on future works.

## 2 Previous works on combining Supervised and Unsupervised Classifiers for Intrusion Detection

In [12], [79] authors combine an unsupervised (namely, clustering) strategy with a supervised classifier: clustering-dependent additional features are provided to the supervised classifier which ultimately decides on the classification. Their results are encouraging, despite they identify a key issue which they describe as "augmenting data sets with too many scores could be detrimental due to overfitting and variance issues" [12]. Briefly, their experimental data showed that adding knowledge derived by unsupervised tasks may constitute noise and disturbing – rather than helping – the supervised learning process. Similarly, in [15] authors processed the KDDCup99 dataset with a stacker that executes an unsupervised probabilistic model before than a supervised rule generator. They conclude that their approach helps discovering a general pattern that a specific group of outlier may have, despite results are valid only for the KDDCup99 dataset, now considered outdated, and analogous experiments were not repeated on recent datasets and attacks.

Other researchers [14] aimed to "detect and isolate malicious flows from the network traffic and further classify them as a specific type of the known malwares, variations of the known malwares or as a completely new malware strain". They first setup a binary classifier to distinguish between normal and anomalous data flows, while another supervised-unsupervised classifier isolates the specific malware classes. Malware is targeted also in [13], where authors define an hybrid malware detector that examines both the permissions and the traffic features to detect malicious activities in the Android OS. Similarly to [12], they first run the clustering procedure and subsequently build a k-Nearest Neighbour (kNN) classifier that performs better than a kNN that solely relies on dataset features. In [14] these layers are used to perform a multi-class identification of the detected malware, while clustering procedures are used in [13] and [14] to pre-filter data before applying supervised classification.

It is worth noticing that mixed supervised-unsupervised learners are being proposed also in domains other than security. For example, they are proposed to support physics domain either to predict solar flares [18] or to improve performance of auto-encoders [19]. Most recently, they are proposed to help classification in healthcare [17] and to complement image processing [16]. Other studies independently execute heterogeneous classifiers in parallel to later combine their results into a unified score according to static rules e.g., voting [36], which was proven beneficial for redundant systems [37] and is currently applied in many domains.

## 3 Meta-Learning to Combine Supervised and Unsupervised Classifiers

Combining supervised and unsupervised classifiers usually implies building a multi-layer classification process where the result of each individual classifier contributes to the final output and may provide additional information to the other classifiers.

### 3.1 Meta-Learning for Model Combination

We first debate on strategies that are currently used to combine ensembles of heterogeneous and homogeneous classifiers. As defined in [74], a meta-learner is a classifier that uses knowledge acquired during base-learning episodes, i.e., *meta-knowledge*, to improve meta-level classification capabilities. More specifically, a *base-learning* process starts by feeding dataset features into one or more learning algorithms to derive one or more models to be used for classification at a base-level. Results of *base-learners* partially build *meta-features* [75] to be provided alongside with other features to the *meta-level classifier*, which computes the result of the whole meta-learner.

Different meta-learners have been proposed through years, recently summarized in [28] and detailed as follows.

- *Bagging* [21] combines base-learners of the same type by submitting bootstrap replicas of the training set. The unified result of the ensemble is derived by the majority of individual results of base-learners.
- *Boosting* [22] builds an ensemble of homogeneous weak learners. Overall capabilities of a weak learner may not be excellent: the underlying idea of boosting is to orchestrate ensembles of weak learners to build a strong meta-learner.
- *Voting (Weighted)* [36] counts opinions coming from an heterogeneous ensemble of individuals, and provide the final decision based on a linear aggregation (sum) of individual responses. Voters decide on a k out of N (kooN) rule as follows: when at least k out of N individuals (k ≤ N) agree on a result, the result is chosen as the result of the ensemble.
- *Stacking* [23] relies upon different algorithms, trained with the exact same training set, as base-learners. Their outputs become model-based meta-features, which are fed to another independent classifier, the meta-level classifier, to deliver a unified result. Differently from bagging and boosting, the final output is not obtained through majority voting: the independent classifier combines individual results according to a general (and possibly non-linear) function.
- *Cascading* [24] stems from boosting by employing heterogeneous weak learners to increase the system complexity. To derive the final result, the data point is sequentially sent through a sequence of classifiers. The first classifier who is confident enough about its classification produces the output of the cascading meta-learner.
- *Delegating* [26] is very similar to cascading, but all base-learners are trained on the exact same portion of the train set. This meta-learner adopts a cautious approach: if a classifier is not confident enough about its result, the final decision is delegated to the next base-learner.
- *Cascade Generalization* [25] performs a sequential composition of heterogeneous classifiers. Instead of a parallel use of classifiers as in Stacking, Cascade Generalization sequentially applies classifiers such that the ($i$+1)-th classifier is a meta-level learner for the $i$-th classifier. In other words, each classifier of the sequence is both a base-learner and a meta-learner. Similarly to Cascading, the first classifier who is confident enough produces the output of Cascade Generalization.

### 3.2 Meta-Learning to Combine Supervised and Unsupervised Classifiers

Meta-learners have been successfully used in the past to improve supervised classification [31], [32] or more recently for unsupervised learning [28]. However, their suitability for combining supervised and unsupervised classifiers is not straightforward and therefore has to be carefully analysed.

Bagging and Boosting ensembles are homogeneous: therefore, they do not allow the simultaneous usage of supervised and unsupervised classifiers. Since the process of training weak learners in Boosting relies on identifying hard-to-classify areas of the training set (assuming train labels are known) we may think of using unsupervised weak base-learners to build Boosting. As described in [28], this solution outperforms most unsupervised classifiers, but still produces a lot more misclassifications with respect to supervised classifiers when dealing with known attacks.

*Cascading*, *Delegating* and *Cascade Generalization* show potential as they embed heterogeneous base-level classifiers, but heavily rely on the concept of confidence of a classifier with respect to a specific class prediction. Suppose that an algorithm outputs probabilities for each of the two {*normal*, *anomaly*} classes and decides by choosing the class that reports on the highest probability: the higher the probability of belonging to the "majority class", the higher the confidence. However, not all the algorithms clearly assign probabilities to classes. Moreover, confidence may depend on the decision function (see Figure 1, V3) used to convert numeric scores into binary labels, and the way it is calculated varies depending on implementations. Consequently, such confidence-based meta-learners may suffer when there is no standardized way to calculate confidence, which will most likely be calculated differently when considering multiple supervised/unsupervised algorithms and decision functions (e.g., Interquartile Range, [57]).

*Stacking* approaches may partially skip the problem of thresholding and confidence because they rely on model-based features as the numeric outputs of base-learners, which are then fed to the meta-level classifier [76], [83]. This is responsible of aggregating them and delivering a unified result, without calculating any confidence. Consequently, the choice of the meta-level classifier has a relevant role in Stacking and therefore has to be planned carefully.

(Weighted) *Voting* deserves a separate discussion. Although it is widely used to manage redundant (and often diverse) components [37], [36], it cannot be easily adapted to vote on the individual decisions of both supervised and unsupervised classifiers. For example, we may consider three supervised algorithms S1, S2, S3 and two unsupervised algorithms U1 and U2 that are run in parallel and their decisions aggregated by voting using a k out of 5 (koo5) rule. How do we set the *k* to define a koo5 rule without biasing the result towards either the supervised or the unsupervised algorithms? When zero-day attacks happen, supervised algorithms are likely to label a data point as "normal" while unsupervised algorithms U1 and U2 have chances to detect the attack and label the data point as "anomalous". Therefore, setting a threshold (e.g., k = 3) without considering who raised the alarm is not feasible. Moreover, weighted voting approaches may raise the amount of i) False Positives, if more weight is assigned to unsupervised algorithms, or ii) False Negatives, as zero day attacks are likely to be misclassified i.e., undetected, if heavier weight is given to supervised classifiers S1, S2 and S3.

## 4 Supervised-Unsupervised Stacking

Summarizing, Stacking turns out as the most promising meta-learning technique for model combination that have the potential to combine supervised and unsupervised classifiers for Intrusion Detection. Therefore, this section details the main characteristics of Stacking and how we plan to instantiate a *Stacker* to combine Supervised and Unsupervised Intrusion Detectors.

### 4.1 On the Choice of Base-Learners

*Stacking* relies on different base-learners, which are classifiers built using the same dataset features and heterogeneous ML algorithms. Outputs of base learners produce model-based meta-features [75], which, alongside with dataset features, are fed to an independent meta-level classifier that delivers the final result. In addition, dataset features should be provided to the meta-learner alongside with model-based features, which describe the system.

It is evident that the choice of meta-features, and consequently of the base-level classifiers to generate model-based features, plays a highly relevant role in defining base-learners for Stacking. The following combinations of supervised – unsupervised classifiers could be employed to define base-learners and to generate model-based features. As motivated in Section 4.1.2, we consider the adoption of unsupervised base learners as the most beneficial approach out of potential alternatives.

#### 4.1.1 Supervised Base-Learners

Supervised base-learners force the adoption of an unsupervised meta-level classifier to combine both approaches. However, it is acknowledged [2], [20], [30] that the models learned by unsupervised algorithms may misclassify known attacks, raising the overall number of misclassifications and skewing the whole meta-learning towards a traditional unsupervised approach.

#### 4.1.2 Unsupervised Base-Learners

Instead, feeding unsupervised model-based features to a supervised meta-level classifier may balance both approaches. The rationale is the following. Supervised classifiers learn a model that links specific feature values or compound conditions to anomalous labels. This makes them weak to unknowns as they may alter dataset features unpredictably. However, if meta-features are generated by unsupervised base-level learners, the output of the supervised meta-level is directly coupled to those model-based features other than dataset features, with potentially higher detection accuracy in case of zero-day attacks.

### 4.1.3 Supervised and Unsupervised Base-Learners

The last option relies in adopting a mixed approach by using both approaches at the base-level. This has the clear advantage of decoupling the choice of the meta-level learner from the choice of base-level learners: regardless of the choice of the meta-level classifier, both approaches are still used in the base-level and therefore the resulting meta-learner embeds both supervised and unsupervised algorithms. However, this setup will provide relevant prominence either to supervised (if the meta-level classifier is supervised), or to unsupervised learning, unavoidably skewing the balance of the two approaches.

## 4.2 Identification of Model-Based Meta Features

### 4.2.1 Numeric and Binary Algorithms Scores

We investigate potential model-based meta-features to be chosen for our Stacker. As already discussed in Figure 1, we assume each unsupervised algorithm *ua* to compute both the numeric score *ua.num* and the binary score *ua.bin* obtained after applying a decision function to *ua.num* score (steps Te2, Te4 of Figure 1).

Both scores are important: *ua.bin* is the final class predicted by the classifier after applying a given decision function, while *ua.num* score is a plain number that is solely assigned by the algorithm, without further processing.

### 4.2.2 Output from Diverse Unsupervised Algorithms

In addition, we believe it is beneficial to simultaneously employ more than a single unsupervised base-learner to generate model-based meta-features for our Stacker. Let UA be a set of *uk* unsupervised algorithms UA = {ua$_1$, ua$_2$, …, ua$_{uk}$}. In total, the UA set generates 2·*uk* meta-features, or rather it generates *ua$_i$.num* and *ua$_i$.bin* features for each algorithm *ua$_i$*, $0 \leq i \leq uk$.

The UA set cannot be defined without information about i) the detection capabilities of individual algorithms and ii) their synergy. Noticeably, past studies show that the usage of diverse components [55] helps avoiding common mode failures [56] and this suggests the adoption of *diverse* algorithms to reduce misclassifications. However, defining two unsupervised algorithms as diverse is not straightforward. Potentially, all algorithms slightly differ in the way they classify data points: nevertheless, different algorithms may rely on the same heuristics. For example, algorithms as ODIN [61], FastABOD [46], LOF [70] and COF [67] embed a k-NN search either to devise their final score or to reduce computational costs. In this case, algorithms differ from each other but are not completely diverse as they all share the concept of neighbours. To such extent, in this study *we consider two unsupervised algorithms diverse if they do not belong to the same family* (i.e., clustering, density, neural network, angle, statistical, classification, neighbour, see Section 1.2).

### 4.2.3 Voting Counters as Meta-Features

Lastly, we present how voting can be used as an additional meta-feature for our Stacker. A (weighted) voter first calculates a (weighted) counter and then applies a threshold that is learned during training to perform binary classification. To build meta-features, computing the counter may suffice as the threshold may be learned during training of the supervised meta-level classifier.

Considering the UA set of algorithms that generate *num* and *bin* meta-features, we define the two counters in (1) and (2).

$$Count(UA) = \sum_{ua_i \in UA} ua_i.bin \quad (1)$$

$$WCount(UA, rep) = \frac{1}{|UA|} \sum_{ua \in UA} \frac{1}{2}[1 - (-1)^{ua.bin} * rep(ua)],$$

$$-1 \leq rep(ua) \leq 1 \quad (2)$$

The counter *Count(UA)* simply represents the amount of algorithms *ua ϵ UA* that have the feature *ua.bin* set to 1, or rather the amount of unsupervised algorithms that consider a given data point as anomalous. Consequently, $0 \leq Count(UA) \leq |UA|$. Instead, the counter *WCount*(*UA, rep*) calculates a weighted count, which extends Count(UA) by adding the concepts of reputation. Each algorithm *ua ϵ UA* is associated to a quantity *rep(ua)* ϵ [-1; 1], where -1 represents an algorithm that is always wrong, 1 identifies an algorithm that always correctly classifies data points, and 0 describes an algorithm which is equivalent to random guessing. This allows calculating the weighted counter *WCount(UA, rep)* that ranges from 0, i.e., very confident that the data point describes a normal behaviour, to 1, which is achieved only when all *ua ϵ UA* have maximum reputation and agree on the anomaly.

Figure 2 wraps up the discussion on meta-features by showing a practical example that considers sample values of UA = {*AlgA, AlgB*}, *rep*(*AlgA*) = 0.2 and *rep*(*AlgB*) = *0.8*.

## 4.3 Selection of the Meta-Level Classifier

The adoption of unsupervised base-learners generates model-based meta-features that can be used by the supervised meta-level classifier to detect attacks. Such classifier should be chosen amongst algorithms that have been proven effective. Supervised binary classifiers are widely used since decades and therefore there are plenty of available solutions. We do not put any constraint on the meta-level classifier and therefore in this work we investigate different families of supervised classifiers, namely neighbour-based [39], boosting [32], decision trees and ensembles [31], support vector machines [33], statistical methods [40] and embedded deep neural networks [59].

Figure 2: Example of Dataset Features (green solid box), Algorithms Features (red dashed box) and voting counters (others).

| packets | bytes | Duration | AlgA.num | AlgB.num | AlgA.bin | AlgB.bin | Count | WCount | label |
|---|---|---|---|---|---|---|---|---|---|
| 13 | 13005 | 0 | 0.02 | 20552 | 1 | 1 | 2 | 0.775 | anomaly |
| 11 | 1292 | 0 | 0.17 | 1811 | 1 | 1 | 2 | 0.775 | anomaly |
| 271 | 13193 | 13 | 0.02 | 20857 | 1 | 1 | 2 | 0.775 | anomaly |
| 1 | 142 | 0 | 0.50 | 29 | 1 | 0 | 1 | 0.375 | normal |
| 1 | 160 | 0 | 4.78 | 0 | 0 | 0 | 0 | 0.225 | normal |

Figure 3: Methodology adopted in this paper. Labels S1 to S5 graphically map the five steps, following Section 5.1.

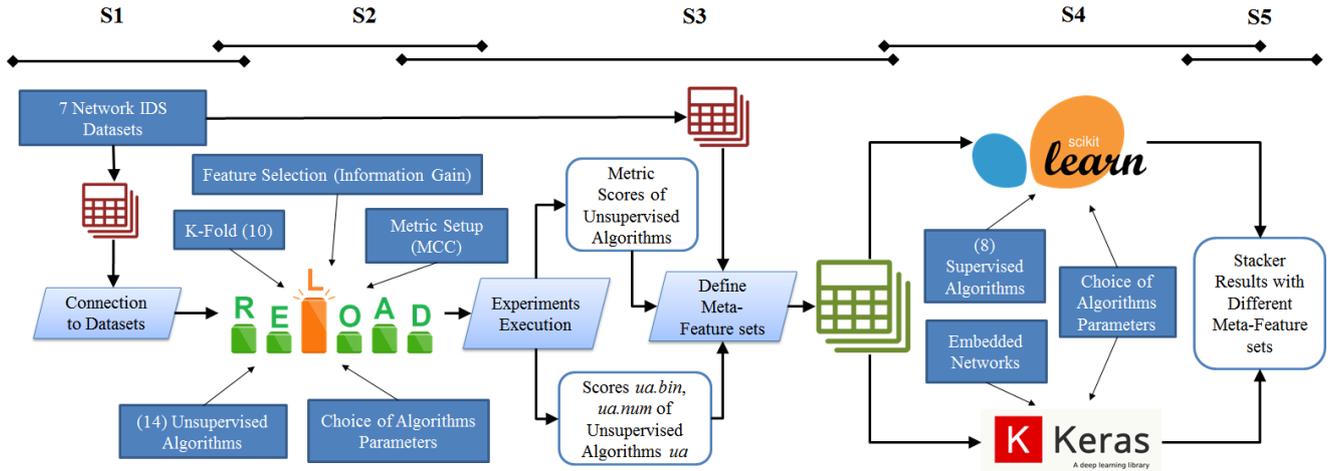

## 5 Experimental Setup

### 5.1 Methodology

To elaborate and evaluate our Stacker, we performed an experimental campaign organized in five steps S1-S5 (Figure 3).

S1. We collect public attack datasets that contain network data of real or simulated systems, creating variants that allow analysing unknown attacks (Section 5.2).

S2. Then, we review the literature to select i) unsupervised algorithms that are suitable for anomaly detection in Section 5.3, and ii) evaluation metrics in Section 5.4.

S3. Next, we apply each unsupervised algorithm on each dataset to devise model-based meta-features and collect metric scores, which will allow identifying meta-feature sets. This is described in Section 5.5.

S4. Afterwards, in Section 5.6 we identify from literature the supervised binary classifiers that we will exercise as meta-level learners on the meta-features computed in the step S3.

S5. Last, the experiment execution is described in Section 5.7.

The execution of experiments of S5 will allow elaborating in Section 6 on results, which will offer a wide variety of insights, showing evidence of the superiority of the stacking approach and identifying the recommended set-up for a supervised-unsupervised Stacker.

### 5.2 The Selected Public Datasets

This section reports the seven public datasets that we used in our study. Table 1 summarizes the datasets reporting name, reference,

Table 1: Datasets used in this study. We report on the name, the size of the subset we selected, number of ordinal features, categories, frequency of attacks, and dataset release year.

| Name | # Data Points | Ordinal Features | Categories of Attacks | % Attacks | Year |
|---|---|---|---|---|---|
| ADFA-Netflow | 132 002 | 3 | 5 | 31.03 | 2017 |
| CIDDS-001 | 400 000 | 7 | 4 | 39.16 | 2017 |
| ISCX12 | 400 000 | 6 | 4 | 7.96 | 2012 |
| NSL-KDD | 148 516 | 37 | 4 | 48.12 | 2009 |
| UGR16 | 207 256 | 7 | 5 | 12.99 | 2016 |
| CICIDS17 | 500 000 | 75 | 4 | 29.88 | 2017 |
| UNSW-NB15 | 175 341 | 38 | 7 | 38.06 | 2015 |

amount of data points, number of different attacks, and percentage of attack records in each dataset.

**Selected Datasets.** Starting from the recent surveys [41], [90] and by querying online portals we selected datasets with the following characteristics: i) published in the last decade, ii) labeled (at least partially), iii) containing at least 100.000 data points to guarantee they contain a sufficient amount of data, and iv) previously used for anomaly-based intrusion detection, to compare our scores with others. Our selection process resulted in the following datasets: ISCX12 (portions of Sunday, Monday, Tuesday, Thursday logs to include 4 categories of attacks [42]), UNSW-NB15 (full train set, [45]), UGR16 (week4 portion, [50]), ADFA-NetflowIDS (full dataset, [51]), CICIDS17 (portions of Tuesday, Wednesday and Friday-Afternoon logs to include 4 categories of attacks, [53]) and CIDDS-001 (files '9' and '20', [44]). Moreover, we also included NSL-KDD (full train and test portions, [43]) as this dataset is often used as benchmark, despite it is more than 10-years-old.

**Datasets with Unknowns.** To enable an analysis of detection capabilities of unknown attacks, we create variants of each dataset $dn\_att$ where i) $dn$, dataset name, is the name of the dataset, and ii) $att$, is one of the categories of attacks contained in the $dn$ dataset. Each dataset variant $dn\_att$ is organized into a 50% train/test split where the train partition does not contain data points related to the attack $att$. Consequently, data points connected to $att$ are unknown during training and are equivalent to zero-day attacks during testing. For example, ISCX12_BruteForce indicates a variant of the ISCX12 dataset where i) the train partition contains normal data and data related to all attacks except BruteForce i.e., DoS, DDoS and Infiltration attacks, and ii) the test partition contains normal data and data related to all the attacks DoS, DDoS, Infiltration and BruteForce. We create a total of 33 variants, one for each attack in each dataset: 5 for ADFA-Netflow, 4 for CIDDS-001, 4 for ISCX12, 4 for NSL-KDD, 5 for UGR16, 4 for CICIDS17 and 7 for UNSW-NB15. Overall, our experiments will use 7 full datasets and 33 variants, for a total of 40 sets.

### 5.3 Unsupervised Classifiers (Base-Level)

We choose a set of unsupervised binary classifiers to perform intrusion detection on the selected datasets. To guarantee diversity of base-learners, we select a pool of algorithms that are as heterogeneous as possible and belong to angle, clustering, neural networks, density-based, neighbour-based, statistical, and classification families. We disregard heavy algorithms (e.g., ABOD

Table 2: Unsupervised algorithms that we use as base-learners to generate model-based meta-features. Parameters and complexities refer to the implementations available in the RELOAD [52] tool.

| Algorithm | Family | Parameters | Complexity (n: train set size) | |
|---|---|---|---|---|
| | | | Train | Test |
| COF | Density, Neighbour | k: number of neighbours | $O(knlog(n))$ | $O(klog(n))$ |
| DBSCAN | Clustering | minPts, eps | $O(nlog(n))$ | $O(1)$ |
| FastABOD | Angle, Neighbour | k: number of neighbours | $O(n^2+nk^2)$ | $O(k^2log(n))$ |
| G-Means | Clustering | (automatic search of cluster number c) | $O(cn)$ | $O(c)$ |
| HBOS | Statistical | b: number of bins | $O(n)$ | $O(1)$ |
| iForest | Classification | t: number of trees, s: tree samples | $O(tslog(s))$ | $O(tlog(s))$ |
| K-Means | Clustering | c: number of clusters | $O(cn)$ | $O(c)$ |
| kNN | Neighbour | k: number of neighbours | $O(nlog(n))$ | $O(klog(n))$ |
| LDCOF | Density, Clustering | c: number of clusters | $O(c^2+cn)$ | $O(c^2)$ |
| LOF | Density, Neighbour | k: number of neighbours | $O(kn+nlog(n))$ | $O(klog(n))$ |
| ODIN | Neighbour | k: number of neighbours | $O(nlog(n))$ | $O(klog(n))$ |
| SDO | Density | obs: number of observers, q | $O(obsn)$ | $O(obs)$ |
| SOM | Neural Network | e:epochs, x: neurons | $O(nex)$ | $O(x)$ |
| 1-Class SVM | Classification | kernel, nu: fraction of margin errors | $O(n^2)$ | $O(1)$ |

[46], which has cubic time complexity), as this study already builds on meta-learning, which naturally requires many computing and memory resources. We select 14 algorithms as follows:

- One algorithm for each family: *One-Class SVM* (classification family, [62]), *K-Means* (clustering, [73]), *kNN* (neighbour, unsupervised variant in [71]), *HBOS* (statistical, [60]), *SOM* (neural-network, [64]), *FastABOD* (Angle-Based, [46]), and *LOF* (density-based, [70]).
- Other well-known algorithms as *DBSCAN [69]*, *COF [67]*, *LDCOF [66]*, *Isolation Forests [72]*, *G-Means [65]*, *ODIN [61]*, *Sparse Density Observers* (SDO, [63]) to widen the selection of unsupervised algorithms.

Table 2 summarizes the 14 algorithms, which have at most quadratic complexity for training and constant test time, except for neighbour-based algorithms which perform kNN search and require at least logarithmic time using K-D trees. In addition, we looked for public frameworks that allow running unsupervised algorithms on datasets. After examining different options, we chose RELOAD [52], a Java open-source tool that wraps unsupervised algorithms from ELKI and WEKA, and includes additional implementations of unsupervised algorithms. The tool also allow to setup grid searches to devise optimal values of algorithms parameters. Besides G-Means, which does not rely on parameters, we try the following combinations of parameters.

- Values *k* for kNN-based algorithms, samples *s* and trees *t* building iForest, number *hist* of histograms in HBOS and observers *obs* of SDO are chosen in the set {1, 2, 3, 5, 10, 20, 50, 100, 200, 500}.

Other algorithms have specific parameters as follows:

- One-class SVM may be created either with {linear, quadratic, cubic, radial basis function} kernels and *nu*, which affects the amount of support vectors to be created, in {0.01, 0.02, 0.05, 0.1, 0.2}.
- In addition to *obs*, SDO needs also a *q* ϵ {0.05, 0.1, 0.2, 0.5} threshold that the algorithm uses to derive the "closest" observers.
- Lastly, DBSCAN clustering uses a combination of the minimum number of data points in a cluster *minPts* ϵ {1, 2, 3, 5, 10} and *eps* ϵ {100, 200, 500, 1000}, which defines the radius of the cluster around each data point.

For each algorithm and each dataset, grid searches discover the parameter values that allow obtaining the best MCC score in a small portion of dataset (not overlapping with the testing set), which is used for validation. These grid searches are automatically managed by RELOAD.

### 5.4 Metrics for Evaluation

The effectiveness of binary classifiers is usually assessed by calculating correct classifications (true positives TP, true negatives TN) and misclassifications (false negatives FN, false positives FP), which build the so-called *confusion matrix*. From [49], commonly used metrics aggregate items of the confusion matrix to build Precision, Recall (or Coverage), False Positive Rate (FPR), Accuracy (ACC), FScore-β (Fβ), F-Measure (F1), Area Under ROC Curve (AUC) and Matthews Coefficient (MCC).

Each metric has its own strengths and weaknesses which must be discussed considering the specific domain. In particular, intrusion detectors should primarily focus in reducing FNs, that is, when attacks are not detected. However, it is also evident that a very suspicious IDS - which heavily reduces the amount of FNs at the price of increasing FPs – may offer satisfying detection capabilities but at the cost of many false alarms. Consequently, in this paper we mainly evaluate detection capabilities of ML algorithms with the following metrics.

- Misclassifications (*Misc*), that is the percentage of misclassifications with respect to the whole dataset, defined as *Misc = 1 – Accuracy*.
- Matthews Coefficient (*MCC*) [48], which aggregates all classes of the confusion matrix and provides results that suit also unbalanced datasets [47] i.e., containing mostly normal data points and only few attacks.
- *Recall* [49], an FN-oriented metric that shows the percentage of detected attacks out of all attacks. Whenever applicable, we also calculate *Recall-Unk*, or rather the recall which is only related to detection of attacks that are not included in the training set (if any).

### 5.5 Meta-Feature Sets

Let *UA = {COF, DBSCAN, FastABOD, G-Means, HBOS, iForest, K-Means, kNN, LDCOF, LOF, ODIN, SDO, SOM, One-Class SVM}* be the set of 14 unsupervised algorithms we presented in Section 5.3, and let *rep* be a 14-items array that contains the MCC ϵ [-1; 1] score that each algorithm obtains on a given dataset. To explore how different feature sets impact misclassifications of the meta-level classifier (and of the stacker as a whole) we derive the feature sets *DataF, MetaF(A)* and *FullF(A)* as follows.

$$DataF = \{\forall f \in Dataset \mid f \text{ contains ordinal values}\}$$
$$MetaF(A) = \{ua.num, ua.bin \; \forall ua \in A\} \cup Count(A)$$
$$\cup \; WCount(A, rep(A))$$
$$FullF(A) = DataF(A) \cup MetaF(A)$$

In a nutshell, DataF is the set of all features of a target dataset. MetaF(A) contain the *ua.num* and *ua.bin* for all the algorithms, plus the voting counters Count(A) and WCount(A,rep(A)). FullF(A) merges DataF and MetaF(A) features.

### 5.6 Supervised Classifiers (Meta-Level)

Supervised classifiers will implement the meta-level learner of the Stacker. Different algorithms may suit the implementation of this layer: we select the well-known K-Nearest Neighbors (kNN, [39]), Support Vector Machines (SVMs, [33]), Decision Trees , Random Forests [31], ADABoost [32], Naive Bayes and Quadratic Discriminant Analysis [40], whose implementations are all made available in the *Scikit-Learn* [1] *Python* package. Algorithms parameters, whose best configuration is selected through grid searches similarly to RELOAD, range as described below.

- *K-NN* with values of k ∈ {1, 3, 10, 30, 100} and Euclidean distance as reference distance function.
- SVM: we individually instantiate three different SVMs using Linear, RBF or Polynomial (Quadratic) kernels.
- Decision Tree: we adopted different depth limits in the range {5, 20, no limit} to define how the tree is built.
- Boosting: AdaBoostM2 algorithm with {10, 30, 100, 200} trees.
- Random Forest: we devised 6 parameter combinations by using {10, 30, 100} trees, either with unlimited depth of decision trees, or with maximum depth limited to 10.
- Naïve Bayes and Quadratic Discriminant Analysis (QDA): usage of default parameters provided by Scikit-Learn.

Further, we consider neural networks enhanced with entity embedding, which improves fitting on tabular data when the data is sparse and/or statistics are unknown. With entity embedding, similar values are mapped in the embedding space to reveal the intrinsic properties of the categorical (discrete) variables [58]. The Python code for entity embedding was crafted starting from [59] and made available at [93] .

### 5.7 Experiments Setup and Execution

We describe here the experimental setup for our study.

**Feature Selection (Base-Learners).** We apply Information Gain [91] to automatically rank and extract the most relevant features out of each dataset, filtering out noisy features.

**Feature Selection (Meta-Level Learner).** We let supervised algorithms to automatically select and derive features (as it is the case of representation learning for deep learners [58]) by providing them an entire meta-feature set.

**Train-Test Split.** We generate model-based unsupervised features for the full dataset (and variants) by adopting a 30-70 train-test split with 10-fold sampling, or rather the biggest training subset that did not make the Java-based RELOAD escalate into memory errors.

For supervised algorithms to be used as meta-level learners, we proceed to a 50-50 train-test split.

**Machine to Execute Experiments.** Once all the parameters mentioned above are set, we run the experimental campaigns including all the datasets and algorithms considered in this study. The experiments have been executed on a server equipped Dell Precision 5820 Tower with a 12-Core I9-9920X and GPU Nvidia Quadro RTX5000, and 2.5 TB of user storage.

**Experiments Execution.** Overall, executing the experiments required approximately 50 days of 24H execution. All the metric scores and files that we used to collect and summarize values are publicly available at [94].

## 6 Comparing the Stacker with Supervised and Unsupervised Classifiers

This section discusses results of our experimental campaign to show the improvements in adopting the Stacker with respect to either supervised or unsupervised ML algorithms to detect intrusions.

### 6.1 Information carried by Unsupervised Meta-Features

First, we aim at understanding if and how unsupervised meta-features can provide additional information that has the potential to reduce misclassifications of the Stacker with respect to traditional classifiers. In Table 3 we report the highest information gain (column "1" in Table 3; the highest value assigned to one of the available features), and the average information gain computed using the top 3, 5, 10 features with the highest information gain (columns "3", "5", "10" respectively). Those aggregated scores of information gain are computed for the DataF, MetaF(UA), and FullF(UA) of each dataset in Table 1.

Table 3: Information Gain scores of the best 1, 3, 5, 10 features of different feature sets for each of the 7 datasets.

| Dataset | Feature Set | # Feat. | Information Gain (Avg) | | | |
|---|---|---|---|---|---|---|
| | | | 1 | 3 | 5 | 10 |
| ADFANet | DataF | 3 | .859 | .381 | .381 | .381 |
| | MetaF(UA) | 30 | .865 | .865 | .864 | .860 |
| | FullF(UA) | 33 | .865 | .865 | .864 | .861 |
| CICIDS17 | DataF | 75 | .491 | .478 | .447 | .407 |
| | MetaF (UA) | 30 | .445 | .440 | .438 | .434 |
| | FullF(UA) | 105 | .491 | .478 | .463 | .449 |
| CIDDS-001 | DataF | 7 | .886 | .708 | .527 | .387 |
| | MetaF (UA) | 30 | .855 | .818 | .788 | .737 |
| | FullF(UA) | 37 | .886 | .855 | .828 | .773 |
| ISCX12 | DataF | 6 | .339 | .317 | .272 | .248 |
| | MetaF (UA) | 30 | .319 | .303 | .248 | .192 |
| | FullF(UA) | 36 | .339 | .331 | .317 | .260 |
| NSL-KDD | DataF | 37 | .746 | .657 | .588 | .500 |
| | MetaF (UA) | 30 | .700 | .671 | .652 | .635 |
| | FullF(UA) | 67 | .746 | .703 | .677 | .648 |
| UGR16 | DataF | 7 | .442 | .285 | .201 | .147 |
| | MetaF (UA) | 30 | .402 | .399 | .394 | .382 |
| | FullF(UA) | 37 | .442 | .413 | .406 | .390 |
| UNSW-NB15 | DataF | 38 | .592 | .520 | .428 | .319 |
| | MetaF (UA) | 30 | .651 | .613 | .595 | .553 |
| | FullF(UA) | 68 | .651 | .615 | .602 | .563 |

---
[1] https://scikit-learn.org/stable/unsupervised_learning.html

For example, we focus on the UNSW-NB15 dataset at the bottom of the table. We observe that average values of information gain when considering FullF(UA) features are higher than DataF, which does not contain model-based meta-features. This trend holds when we consider the quantity of information carried by the best feature, that is the column labelled as "1" in Table 3. The value assigned to FullF(UA) is higher than DataF and overlaps with MetaF(UA): this means that the most informative feature is a meta-feature (namely, feature WCount(UA)). Consequently, we can expect that meta-level classifiers will build upon such feature when building their models and therefore it will contribute to improve classification. A similar behaviour can be observed for ADFANet dataset (on top of Table 3), which indeed has different characteristics with respect to UNSW as it contains only 3 ordinal dataset features. This has multiple implications: i) the average of the best 3, 5, 10 features of DataF will be the same, and ii) the addition of meta-features will likely provide useful additional information since the number of feature dramatically raises from 3 to 33 (see # Feat. column in Table 3).

Overall, the information carried by FullF(UA) features is (significantly) higher than DataF in all datasets when looking at 5 and 10 features, and often improves the score of the best feature or the top three features. This analysis quantitatively shows that meta-features carry relevant information and therefore have potential to enhance classification.

Table 4: Misclassifications, MCC, Recall, Recall_Unk (recall on unknowns-only) achieved by algorithms on each of the 7 dataset and their variants. For each metric, we report scores obtained by using Supervised algorithms with DataF features, Stacker with MetaF(UA) and FullF(UA) features, and Unsupervised Algorithms using DataF features.

| Dataset | Unknown Attacks (Variant) | % Unkn | Misclassifications (%) | | | | MCC | | | | Recall | | | | Recall-Unk | | |
|---|---|---|---|---|---|---|---|---|---|---|---|---|---|---|---|---|---|
| | | | Sup | Stacker | | Unsup | Sup | Stacker | | Unsup | Sup | Stacker | | Unsup | Sup | Stacker | |
| | | | DataF | MetaF(UA) | FullF(UA) | DataF | DataF | MetaF(UA) | FullF(UA) | DataF | DataF | MetaF(UA) | FullF(UA) | DataF | DataF | MetaF(UA) | FullF(UA) |
| ADFANet | - | 0.0 | 1.33 | 0.32 | 0.44 | 2.41 | .969 | .992 | .990 | .945 | .996 | .998 | .999 | .984 | - | - | - |
| | Mailbomb | 0.8 | 1.96 | 0.32 | 0.44 | 1.33 | .956 | .992 | .990 | .970 | .996 | .998 | .999 | .997 | 1.000 | 1.000 | 1.000 |
| | Other | 2.7 | 1.96 | 1.17 | 1.17 | 1.33 | .956 | .973 | .973 | .970 | .980 | .980 | .972 | .997 | .750 | .776 | .776 |
| | Portsweep | 3.8 | 1.96 | 0.32 | 0.44 | 0.70 | .956 | .992 | .990 | .984 | .996 | .998 | .998 | 1.000 | 1.000 | 1.000 | 1.000 |
| | 1b | 10.3 | 0.47 | 0.95 | 0.60 | 0.67 | .989 | .978 | .986 | .984 | .991 | .986 | .964 | 1.000 | .980 | .974 | .960 |
| | Neptune | 13.4 | 1.96 | 0.32 | 0.44 | 0.70 | .956 | .992 | .990 | .984 | .996 | .998 | .999 | 1.000 | 1.000 | 1.000 | 1.000 |
| CICIDS17 | - | 0.0 | 0.03 | 0.10 | 0.03 | 1.85 | .999 | .959 | .999 | .915 | .959 | .998 | .998 | .861 | - | - | - |
| | Patator | 1.0 | 0.76 | 1.60 | 1.02 | 1.90 | .966 | .927 | .954 | .912 | .878 | .921 | .941 | .857 | .295 | .295 | .295 |
| | PortScan | 11.2 | 11.14 | 6.81 | 11.12 | 3.66 | .346 | .722 | .349 | .923 | .862 | .138 | .138 | .958 | .007 | .992 | .806 |
| | DoS | 17.7 | 0.75 | 1.00 | 0.75 | 2.41 | .966 | .955 | .966 | .888 | .935 | .942 | .942 | .858 | .000 | .000 | .000 |
| CIDDS-001 | - | 0.0 | 0.97 | 1.27 | 0.19 | 12.15 | .980 | .974 | .996 | .746 | .998 | .998 | .997 | .753 | - | - | - |
| | BruteForce | 0.1 | 0.82 | 1.27 | 1.27 | 12.31 | .983 | .974 | .974 | .746 | .997 | .997 | .997 | .876 | .000 | .126 | .077 |
| | PingScan | 0.2 | 0.84 | 1.28 | 0.21 | 13.74 | .982 | .974 | .996 | .713 | .997 | .997 | .997 | .837 | .978 | .988 | .963 |
| | PortScan | 9.4 | 9.15 | 9.41 | 3.35 | 11.89 | .815 | .810 | .931 | .758 | .762 | .919 | .763 | .718 | .060 | .661 | .673 |
| | DoS | 29.5 | 29.31 | 30.73 | 25.00 | 6.55 | .405 | .356 | .489 | .726 | .245 | .392 | .535 | .821 | .025 | .318 | .195 |
| ISCX12 | - | 0.0 | 3.55 | 3.21 | 1.20 | 5.94 | .796 | .761 | .916 | .489 | .670 | .903 | .899 | .290 | - | - | - |
| | DoS | 0.6 | 3.03 | 3.22 | 1.67 | 6.64 | .803 | .758 | .882 | .464 | .641 | .830 | .861 | .386 | .004 | .392 | .214 |
| | BruteForce | 0.9 | 3.85 | 3.12 | 1.45 | 6.54 | .766 | .768 | .897 | .484 | .686 | .838 | .702 | .421 | .001 | .222 | .270 |
| | DDoS | 3.1 | 3.86 | 5.76 | 3.77 | 6.00 | .702 | .524 | .711 | .480 | .378 | .556 | .538 | .263 | .000 | .231 | .322 |
| | Infiltration | 3.4 | 6.43 | 3.21 | 3.27 | 5.52 | .536 | .761 | .759 | .544 | .673 | .685 | .743 | .385 | .530 | .538 | .539 |
| NSL-KDD | - | 0.0 | 0.55 | 1.10 | 0.61 | 9.07 | .989 | .978 | .988 | .817 | .993 | .997 | .996 | .877 | - | - | - |
| | u2r | 0.1 | 0.58 | 1.13 | 0.63 | 9.89 | .988 | .977 | .987 | .801 | .993 | .996 | .997 | .876 | .267 | .400 | .550 |
| | r2l | 2.8 | 2.57 | 2.87 | 2.46 | 9.36 | .949 | .942 | .951 | .812 | .954 | .959 | .963 | .868 | .056 | .129 | .155 |
| | Probe | 9.5 | 4.44 | 4.49 | 4.88 | 10.94 | .911 | .910 | .902 | .780 | .923 | .923 | .919 | .853 | .282 | .604 | .651 |
| | DoS | 12.4 | 5.03 | 5.34 | 4.95 | 8.86 | .901 | .893 | .903 | .824 | .836 | .889 | .907 | .850 | .866 | .896 | .865 |
| UGR16 | - | 0.0 | 0.44 | 2.19 | 0.33 | 3.98 | .980 | .907 | .985 | .759 | .953 | .975 | .977 | .808 | - | - | - |
| | BlackList | 0.4 | 0.71 | 2.47 | 0.73 | 7.01 | .969 | .893 | .968 | .708 | .925 | .944 | .931 | .797 | .000 | .040 | .000 |
| | NerisBotnet | 0.4 | 0.65 | 2.88 | 0.66 | 6.21 | .971 | .872 | .971 | .734 | .880 | .886 | .952 | .798 | .015 | .000 | .000 |
| | Anom-Spam | 0.7 | 1.08 | 2.54 | 1.14 | 5.15 | .952 | .889 | .949 | .771 | .915 | .912 | .914 | .796 | .000 | .000 | .000 |
| | DoS | 2.3 | 1.64 | 3.24 | 0.25 | 6.38 | .926 | .856 | .989 | .733 | .867 | .982 | .974 | .815 | .514 | 1.000 | 1.000 |
| | Scan44 | 9.2 | 6.01 | 4.93 | 4.42 | 6.74 | .709 | .765 | .792 | .720 | .641 | .715 | .880 | .809 | .380 | .788 | .702 |
| UNSW-NB15 | - | 0.0 | 6.76 | 5.95 | 2.81 | 17.31 | .862 | .879 | .941 | .653 | .971 | .982 | .981 | .870 | - | - | - |
| | Worms | 0.1 | 6.73 | 6.01 | 3.04 | 17.57 | .862 | .878 | .937 | .626 | .971 | .984 | .981 | .763 | .985 | 1.000 | 1.000 |
| | Shellcode | 0.7 | 6.81 | 5.89 | 2.82 | 17.77 | .862 | .880 | .941 | .622 | .970 | .981 | .980 | .757 | .984 | .956 | .984 |
| | Backdoor | 1.1 | 6.15 | 5.84 | 2.95 | 17.43 | .873 | .881 | .939 | .630 | .970 | .984 | .981 | .766 | 1.000 | .993 | .999 |
| | Analysis | 1.2 | 6.10 | 5.93 | 2.90 | 17.76 | .874 | .879 | .939 | .622 | .968 | .981 | .973 | .758 | .833 | .985 | .907 |
| | Reconnaiss. | 6.3 | 6.58 | 6.14 | 2.93 | 17.77 | .866 | .874 | .939 | .622 | .964 | .982 | .980 | .757 | .977 | .968 | .996 |
| | DoS | 7.4 | 7.46 | 6.02 | 3.67 | 17.71 | .848 | .877 | .924 | .623 | .967 | .982 | .980 | .757 | .967 | .988 | .995 |
| | Exploits | 10.3 | 5.76 | 5.55 | 2.81 | 17.42 | .879 | .886 | .941 | .630 | .968 | .978 | .976 | .766 | .968 | .992 | .990 |
| | Fuzzers | 11.0 | 10.37 | 7.57 | 9.08 | 16.27 | .784 | .839 | .811 | .659 | .855 | .782 | .781 | .811 | .198 | .561 | .272 |
| **Averages** | | | **4.26** | **4.09** | **2.80** | **8.62** | **.870** | **.870** | **.911** | **.746** | **.862** | **.913** | **.900** | **.785** | **.482** | **.631** | **.611** |

## 6.2 Discussion and Comparisons on Detection of Attacks

Then, we compare and discuss classification capabilities of the Stacker and classic supervised and unsupervised algorithms with the aid of Table 4, which provides a view on Misclassifications, MCC, Recall and Recall-Unk metrics.

For each metric, the table reports different columns, namely: i) Sup - DataF, that describes scores achieved by supervised algorithms in Section 5.6 on each dataset by using DataF features, ii) Stacker - MetaF(UA), the Stacker results by using MetaF(UA) features, iii) Stacker - FullF(UA), the Stacker results by using FullF(UA) features, and iv) Unsup - DataF, scores of unsupervised algorithms in Section 5.3 using DataF features. The table reports metric scores of Sup - DataF, Stacker – MetaF(UA), Stacker FullF(UA) and Unsup - DataF classifiers that achieved the highest MCC in each dataset or its variants. Recall-Unk is not computed for Unsup – DataF classifiers, because unsupervised algorithms do not distinguish between known and unknown attacks and therefore it is not meaningful to distinguish Recall from Recall-Unk metric.

Low misclassification (Misc) scores point to datasets or variants containing data points that are mostly classified correctly by algorithms: the full dataset and variants of ADFANet, ISCX12, and UGR16 result in less than 5% of misclassifications, as shown in the Misclassification columns of Table 4. On the other hand, the variant CIDDS_DoS scores a very high percentage of misclassifications, which is around 30% when using Supervised (Sup-DataF) algorithms and the Stacker, either using MetaF(UA) or FullF(UA) features. Additionally, misclassifications in CICIDS17_PortScan and UNSW_Fuzzers hover around 10%: on average, a data point out of 10 is going to be misclassified.

Misclassifications of Unsupervised (Unsup-DataF) algorithms do not fluctuate much for the different variants of the same dataset, but they are generally higher than Supervised (Sup-DataF) algorithms and Stacker counterparts. In fact, training goes unlabelled and therefore the addition of data points connected to specific attacks has less impact in defining the normal behaviour than when training supervised classifiers.

These trends become even clearer when considering averages, in the last row of Table 4: supervised algorithms average 4.26% of misclassifications in datasets and their variants, while unsupervised algorithms reach an average of 8.62%. Noticeably, the Stacker-FullF(UA) averages 2.80% of misclassifications, with a 33% of reduction with respect to supervised classifiers. The trend is confirmed also by MCC scores in the centre of Table 4: unsupervised algorithms achieve the lowest average of 0.746, while Stacker-FullF(UA) outperforms all other classifiers reaching an MCC of 0.911. Note that the higher the MCC, the better.

Advantages in adopting the Stacker-FullF(UA) over Sup-DataF is depicted in Figure 4, which shows the difference of MCC score when applying those two classifiers. The adoption of the Stacker is beneficial as it mostly raises MCC i.e., almost all points in Figure 4 are above 0. Moreover, in some cases the growth is noticeable, either because there is a huge difference (e.g., for ISCX_Infiltration the MCC of Stacker with FullF(UA) grows up to 0.759, while supervised achieved only 0.536), or because even a smaller difference is extremely meaningful. In particular, in the full CIDDS dataset Supervised classifiers Sup-DataF achieve a 0.98 of MCC with 1% of misclassifications, while Stacker-FullF(UA) obtains a slightly higher MCC of 0.996, but with only 0.2% misclassifications. Summarizing, Stacker - FullF(UA) shows the highest MCC in 22 datasets or variants out of the 40 in Table 4, and outperforms Sup-DataF in 30 datasets or variants (75% of the cases considered in this study).

## 6.3 Stacker and Zero-Day Attacks

We focus now on undetected attacks, measured as False Negatives and by means of aggregated metrics such as Recall. Recall scores in Table 4 follow a trend similar to MCC: Unsup-DataF show lower scores than Sup-DataF classifiers; Stacker overall improves Recall, again showing the best average scores. With respect to the detection of zero-day attacks, Recall-Unk scores allow elaborating the following aspects:

- Values of Recall and Recall-Unk for Supervised classifiers i.e., Sup-DataF columns in Table 4, are clearly different, confirming that those algorithms lack in detecting zero-day attacks.
- Recall values of unsupervised algorithms i.e., Unsup-DataF in Table 4, are remarkably lower than Recall values achieved by supervised classifiers and by the Stacker, but are clearly higher than their Recall-Unk values. This confirms that unsupervised algorithms generate several misclassifications, but also less FNs when dealing with zero-day attacks.
- Similarly to what happened with Misc and MCC metrics, adopting the Stacker - especially using FullF(UA) features - improves both Recall and Recall-Unk with respect to Sup-DataF supervised classifiers. Figure 5 depicts a scatterplot of the difference of Recall-Unk score against the percentage of unknowns in datasets or variants: this shows how difference is almost always positive, motivating even more the adoption of the Stacker approach. Noticeably, the figure has a trend similar to Figure 4, but absolute difference is in most cases higher (up to 0.8) with respect to the plot regarding MCC.

## 6.4 Supervised Classifiers as Meta-Level Learners

It turns out evident that the adoption of Stacker with FullF(UA) features maximises classification performance for intrusion detection. However, we still need to investigate on which supervised classifier should be used as meta-level learner when instantiating such Stacker. Therefore, Table 5 reports Average and Std of MCC scores of all supervised classifiers to be potentially used as meta-level learners for the Stacker, with different parameters' values. In addition, the table reports on # Best, the number of datasets or variants in which a supervised algorithm achieves the highest MCC.

Figure 4: Difference of MCC achieved by Stacker-FullF(UA) and Sup-DataF plotted against % of unknowns in the test set.

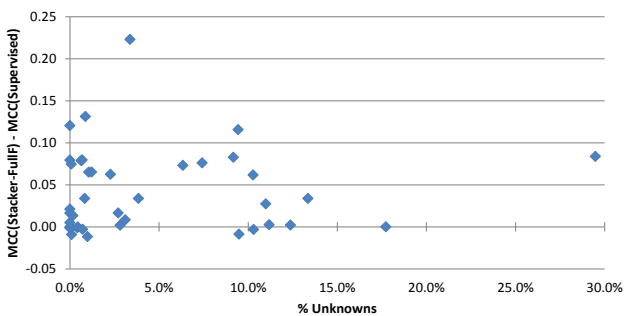

Figure 5: Difference of Recall-Unk achieved by Stacker-FullF(UA) against Sup-DataF plotted against % of unknowns in the test set.

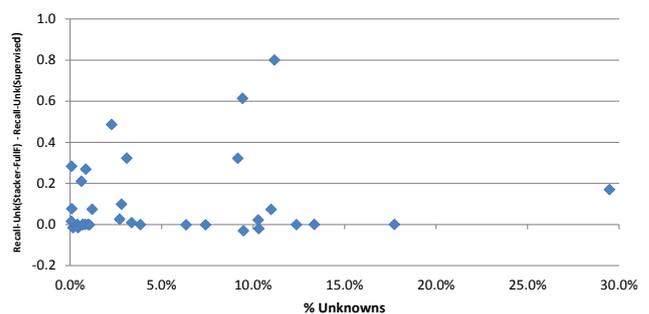

Table 5: Supervised Classifiers to be used as Meta-Level learners, coupled with their parameters values. For each couple, we report the average and standard MCC they achieve on datasets and variants, as well as the times in which an algorithm achieved the maximum MCC, being the preferred choice.

| Meta-Level Algorithm | Parameters | Average MCC | Std MCC | # Best |
|---|---|---|---|---|
| ADABoost | 10 trees | 0.883 | 0.081 | 1 |
| | 30 trees | 0.886 | 0.081 | 1 |
| | 100 trees | 0.795 | 0.276 | 0 |
| | 200 trees | 0.764 | 0.325 | 1 |
| kNN | 1-NN | 0.906 | 0.077 | 1 |
| | 3-NN | 0.902 | 0.081 | 0 |
| | 10-NN | 0.899 | 0.079 | 5 |
| | 30-NN | 0.903 | 0.084 | 0 |
| | 100-NN | 0.897 | 0.093 | 3 |
| Decision Tree | no max depth | 0.750 | 0.211 | 0 |
| | depth: 20 | 0.751 | 0.207 | 2 |
| | depth: 5 | 0.801 | 0.171 | 0 |
| MLP | Default | 0.815 | 0.189 | 0 |
| Embedded Networks | 3 dense layers, batch norm., dropout 0.5 | 0.899 | 0.037 | 10 |
| NaïveBayes | default | 0.674 | 0.312 | 0 |
| QDA | default | 0.538 | 0.472 | 0 |
| SVM | Linear kernel | 0.844 | 0.092 | 0 |
| | Quadratic kernel | 0.866 | 0.097 | 0 |
| | RBF kernel | 0.853 | 0.095 | 0 |
| Random Forest | 10 trees, no depth limit | 0.904 | 0.062 | 0 |
| | 10 trees, depth: 10 | 0.912 | 0.053 | 2 |
| | 30 trees, no depth limit | 0.915 | 0.056 | 1 |
| | 30 trees, depth: 10 | 0.915 | 0.056 | 1 |
| | 100 trees, no depth limit | 0.915 | 0.061 | 12 |
| | 100 trees, depth: 10 | 0.912 | 0.066 | 0 |

Embedded deep neural networks and Random Forests (100 trees, no depth limit) are generally the preferred choice. In particular, Random Forests outperform other supervised algorithms in 16 out of 40 datasets or variants, and it also has the highest average MCC. Multiple instantiations of kNN (1-NN, 3-NN, 30-NN) have average MCC higher than 0.90, but the variability of their scores is higher than Random Forests'. The smallest variability of MCC scores (i.e., lowest Std MCC in the table) belongs indeed to Embedded Networks, which therefore can be considered an overall balanced choice. Statistical methods as Naïve Bayes and QDA show very high variability and therefore are not reliable enough to be used as meta-level learner, while ADABoost and MLP produce many misclassifications.

Wrapping up, *we propose either Random Forests (100 trees) or Embedded Networks as meta-level learners for the Stacker*, as they both show good scores.

## 7 A Stacker for Intrusion Detection

With the help of Figure 6, we instantiate a Stacker that uses FullF(UA) meta-features as derived in this study.

### 7.1 The Overall Architecture

The process of detecting intrusions starts from the target system (i.e., *step* ① in the top left of Figure 6). Here, probes monitor network data, which is logged to build the set of DataF features ②. Those features partially build the FullF(UA) feature set alongside with MetaF(UA) features, which have to be calculated by feeding UA base-learners with DataF features. The process of exercising base-learners can be break down into three partitions.

- *K-Means based base-level classifiers* (③ in Figure 6). K-Means assigns *num* and *bin* scores to a data point after identifying the nearest cluster; this information partially builds LDCOF score, which can therefore be calculated as extension of K-Means scores.
- *kNN-based Base-Level Classifiers* ④. ODIN, FastABOD, kNN, LOF and COF all need to derive the k-nearest neighbours to a given data point and then incrementally compute their score depending on this information. Therefore, the kNN search can be executed only once, delivering its result to those algorithms that can quickly compute the final *num* and *bin* features.
- *Other Unsupervised Base-Level Classifiers* ⑤. Those base-learners assign anomaly scores according to non-overlapping heuristics and therefore have to be exercised independently.

Figure 6: Instantiation of the proposed Stacker. The figure identifies 7 steps ① to ⑦, described in Section 7.

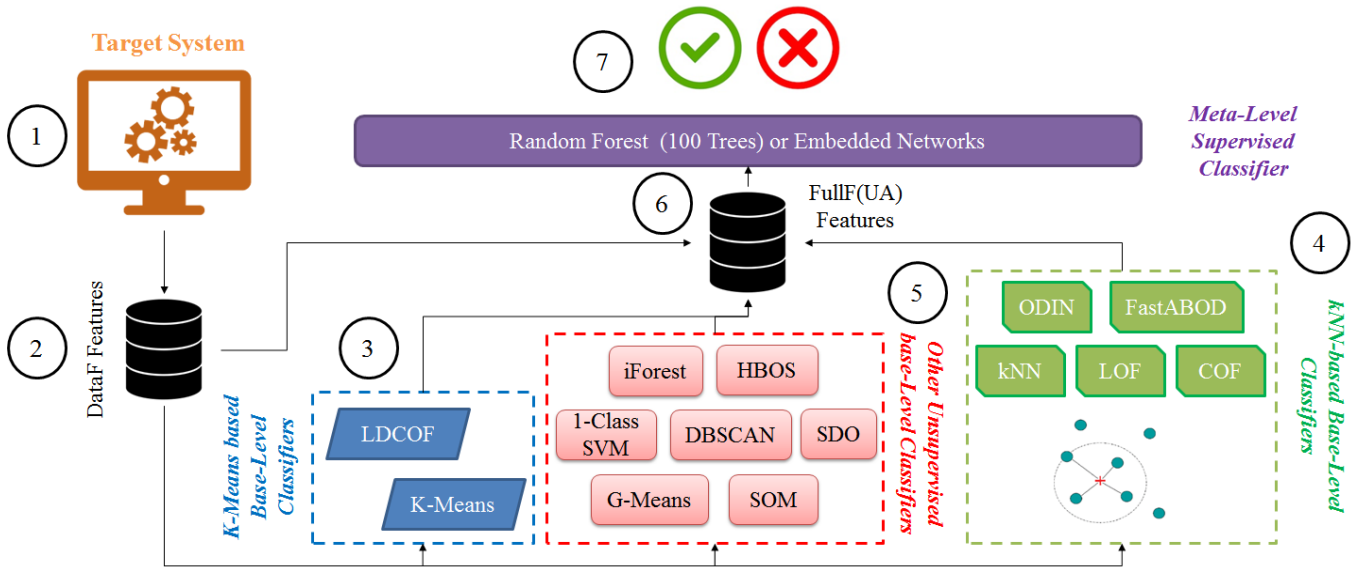

Alongside with DataF features, model-based meta-features *ua.num, ua.bin* generated by unsupervised base-learners *ua ϵ UA* in steps ③, ④, ⑤ are then used to build the FullF(UA) feature set (⑥ in the figure), which is delivered to the supervised meta-level classifier. Such classifier, either Random Forests or Embedded Networks, processes FullF(UA) features and finally decides on anomalies (i.e., step ⑦), completing the precise instantiation of the Stacker(UA).

### 7.2 On the Implementation of the Stacker

Grouping base-learners into ③, ④, ⑤ allows simplifying the overall architecture and avoiding repeating duplicate actions as kNN search. Even after grouping base-learners to avoid duplicate computations, the set of base-learners still requires the implementation of multiple unsupervised algorithms. Those algorithms are mostly well-known and therefore public implementations are available in different programming languages that suit their execution in different environments. Still, at the end of the paper we plan future works to instantiate the Stacker with a set of base-learners that contain less algorithms than UA.

### 7.3 Computational Complexity

Lastly, we elaborate on the computational complexity of the Stacker. The Stacker decides on intrusions by first exercising base-learners and then the meta-level learner; while base-learners may be executed in parallel, the execution of the meta-level learner is necessarily sequential to them. Considering our UA set of 14 unsupervised algorithms, Random Forests as a meta-level learner and a function *cc* that calculates the computational complexity of an algorithm, we obtain that:

$cc(Stacker(FullF(UA))) = max\{cc(ua) \mid ua \in UA\} +$

$cc(Random\ Forests) \approx cc(FastABOD) + cc(Random\ Forests)$

In fact, as it can be noticed in Table 2, FastABOD is the unsupervised algorithm that has the highest asymptotic computational complexity both for train and test phases. Instead, the computational complexity of Random Forests mimics those of iForest, as the time needed to train and test a decision tree and an isolation tree are asymptotically the same. Considering *n* as the size of the training set, we can further elaborate the complexity formula by splitting into train and test complexity as it is shown below.

Train: $cc(Stacker(UA)) \approx O(n^2+nk^2) + O(tslog(s)) \in O(n^2)$

Test: $cc(Stacker(UA)) \approx O(k^2log(n)) + O(tlog(s)) \in O(log(n))$

Aside from *n*, which represent the size of the train set, *k, t, s* are constant algorithm-dependent parameters. Value *k* is obtained through grid searches when training FastABOD, *t* is the number of trees building the Random Forest, which we set to 100 as resulting from Section 6, and *s* is the sample size used to build trees in the Random Forest. Even if we consider *s = n*, the asymptotic complexity of the Stacker is similar to those of many classifiers, which need quadratic time for training and logarithmic time for testing.

Table 6: Studies applying ML-based intrusion detection strategies on public datasets used in this study. Grayed lines point to metric scores achieved by the Stacker described in this paper, while bold-font rows highlight the best metric scores achieved for datasets.

| Study | Dataset | ACC | P | R | F1 | AUC | FPR |
|---|---|---|---|---|---|---|---|
| [87] | | 0.9759 | | 1.000 | 0.983 | | |
| [88] | ADFANet | 0.9530 | 0.995 | 0.958 | | | |
| **Stacker** | | **0.9956** | **0.998** | **0.988** | **0.993** | **0.993** | **0.012** |
| [80] | | 0.9988 | | | | | 0.002 |
| [83] | CICIDS17 | 0.9900 | | | | 1.000 | |
| [82] | | | | | 0.787 | 0.781 | |
| **Stacker** | | **0.9997** | **0.998** | **1.000** | **0.999** | **0.999** | **0.000** |
| [89] | | 0.9954 | | | | 0.969 | |
| [84] | CIDDS-001 | 0.9705 | | | | | 0.021 |
| [82] | | | | | 0.675 | 0.720 | |
| **Stacker** | | **0.9981** | **0.998** | **0.998** | **0.998** | **0.998** | **0.002** |
| **[79]** | ISCX12 | **0.9992** | | **0.989** | | | **0.038** |
| [81] | | 0.9988 | | | | | 0.035 |
| Stacker | | 0.9880 | 0.903 | 0.943 | 0.923 | 0.923 | 0.057 |
| [78] | | 0.9899 | | 0.996 | | | 0.056 |
| [81] | NSL-KDD | 0.9823 | | | | | 0.033 |
| [83] | | 0.9750 | 0.976 | 0.984 | | | 0.075 |
| **Stacker** | | **0.9939** | **0.997** | **0.990** | **0.993** | **0.993** | **0.010** |
| [85] | | 0.9961 | 0.996 | | | | |
| [76] | UGR16 | 0.9719 | 0.988 | 0.981 | | | |
| [86] | | | 0.897 | 0.887 | 0.888 | 0.868 | |
| **Stacker** | | **0.9967** | **0.975** | **0.999** | **0.987** | **0.987** | **0.001** |
| [77] | | 0.9613 | | | | | |
| [76] | | 0.9400 | 0.960 | 0.930 | 0.950 | | |
| [81] | UNSW-NB15 | 0.9371 | | | | | 0.046 |
| [83] | | 0.8800 | | | | 0.860 | |
| **Stacker** | | **0.9719** | **0.982** | **0.946** | **0.964** | **0.964** | **0.054** |

## 8 Comparison with Literature Studies about Intrusion Detection

Ultimately, we position our work with respect to existing studies in the literature that targeted intrusion detection with the datasets we used in this study. Such comparison will not elaborate on the variants of datasets we introduced in this paper, as they are newly created for this work and are not used in any existing study. Instead, we compare metric scores achieved by the Stacker on full datasets with studies we selected according to the following criteria:
- *Recent* (i.e., published in the last 5 years), to guarantee that algorithms are implemented with recent strategies to enhance classification.
- *At least two studies* for each of the 7 datasets we used in this study, to provide a minimum baseline for comparison.
- *Priority* to papers published in international journals rather than conferences, whenever possible.

Our review ended up selecting 15 papers: 2 for ADFANet [87], [88], 3 for CICIDS17 [80], [82], [83], 3 for CIDDS-001 [82], [84], [89], 2 for ISCX12 [79], [81], 3 for NSL-KDD [78], [81], [83], 3 for UGR16 [76], [85], [86], and 4 for UNSW-NB15 [76], [77], [81], [83]. Those papers use metrics different than MCC to evaluate intrusion detectors; consequently, we calculated common metrics achieved by the Stacker on each dataset to enable comparisons. As a result, Table 6 reports on Accuracy (ACC), Precision (P), Recall (R), F-Measure (F1), Area Under ROC Curve (AUC) and False

Positive Rate (FPR) achieved by the studies above and by the Stacker on each dataset. The table is structured as follows: we grouped metric scores of studies targeting the same dataset: for each dataset, we report a greyed line which contains the metric scores achieved by the Stacker. Empty cells in the table happen when related papers do not report on a given metric. The metrics reported in the table were chosen amongst the most commonly used in the papers we selected, and provide a good baseline to compare intrusion detectors.

At a first glance, we may notice how Table 6 reports on many ACC scores, which are provided by all studies but [82], [86] and therefore represent a first term of comparison. In all datasets but ISCX12 the Stacker achieves higher accuracy than literature studies, pointing to a lower amount of misclassifications, either FPs or FNs. This is remarkable as *in our study we do not conduct dataset-specific analyses* (e.g., hand-crafting custom features, pre-processing) that may lead to optimize metric scores for a given dataset. Therefore, the supervised-unsupervised Stacker we built in this paper generally improves accuracy and therefore has a broad range of applicability. However, it is necessary to analyse the case of ISCX12 dataset, where metric scores of the Stacker are slightly lower than its competitors [79], [81]. First, we notice that both studies [79], [81] conduct a cluster analysis to better characterize different attacks, and that works as first step of their analysis and provides them a key advantage as it reveals very effective in increasing metric scores. Moreover, in [79] authors used larger train sets and crafted 30 additional dataset-specific features to help the process, while in [81] authors paired extra-tree classifiers with time windows.

Overall, we can conclude that *the Stacker presented in this paper proved to be very competitive with respect to recent studies, and improved classification scores of known attacks in 6 out of the 7 datasets* we considered. It is also *shaped to be more robust to the detection of unknowns*, which is a very challenging topic for both academia and industry.

## 9   Concluding Remarks and Lessons Learned

This paper motivated the need to combine supervised and unsupervised machine learning algorithms to make network intrusion detectors able to deal with known and unknown (zero-day) threats. To such extent, we analysed available supervised and unsupervised machine learning algorithms, as well as meta-learning approaches which allows combining different classifiers. This process led building a Stacking ensemble, that is, a two-layer meta-learner that orchestrates unsupervised base-level learners and a supervised meta-level classifier. With respect to traditional supervised intrusion detectors, our Stacker reduces misclassifications approximately by 33% and noticeably enhances the detection of zero-day attacks (i.e., Recall on unknown attacks clearly improves). Therefore, it has potential to be adopted as a unified solution for Intrusion Detection. We confirmed the efficacy of the Stacker with an extensive comparison against recent studies which performed intrusion detection on the same public datasets we used; in 6 out of 7 datasets, our Stacker outperformed existing studies.

Our current and future works are primarily directed to investigate on the reduction of the number of unsupervised base-learners to build such Stacker. The objective is to minimize the number of base-learners that are required to make the Stacker easier to implement, but without reducing metric scores. The key argument is that the extent base learners contribute to the decision process varies from learner to learner. We are planning an extensive sensitivity analyses to identify the minimal set(s) of base-learners which have the highest role in the decision process, and guarantee excellent detection capabilities. We are investigating correlation of scores assigned by unsupervised algorithms, to elaborate and quantify their synergy and diversity, rather than using a selection strategy based on qualitative categories as the families of algorithms.